 \documentclass[final,1p,times]{elsarticle}
\usepackage{multirow}
\usepackage{booktabs}
\usepackage{xcolor}

\newcommand{\ug}{{\mathbf{u}}}
\newcommand{\bg}{{\mathbf{b}}}
\newcommand{\Bg}{{\mathbf{B}}}
\newcommand{\Ag}{{\mathbf{A}}}
\newcommand{\omegag}{\mbox{\boldmath${\omega}$}}
\newcommand{\eg}{{\mathbf{e}}}

\usepackage{amssymb}
\usepackage{graphicx}
\graphicspath{{figure/}}
\journal{European Journal of Mechanics B/Fluids}

\begin{document}

\begin{frontmatter}

\title{Cross and magnetic helicity in the outer heliosphere from Voyager 2 observations.}

\author[polito]{M. Iovieno\corref{cor1}}
 \ead{michele.iovieno@polito.it}
 \cortext[cor1]{Corresponding author}
\author[polito]{L. Gallana}
\author[polito]{F. Fraternale}
\author[mit]{J. D. Richardson}
\author[uboston]{M. Opher}
\author[polito]{D. Tordella}
\address[polito]{Dipartimento di Ingegneria Meccanica e Aerospaziale, Politecnico di Torino, Corso Duca degli Abruzzi 24, 10129 Torino, Italy}
\address[mit]{Kavli Institute for Astrophysics and Space Research, Massachusetts Institute of Technology, Cambridge, MA 02139, USA}
\address[uboston]{Department of Astronomy, Boston University, 725 Commonwealth Ave, Boston, MA 02215, USA}


% \title{Cross-helicity in solar wind at 5 AU from Voyager data}


\author{}

\address{}

\begin{abstract}
Plasma velocity and magnetic field measurements from the Voyager 2 mission are used to study solar wind turbulence in the slow solar wind at two different heliocentric distances, 5 and 29 astronomical units, sufficiently far apart to provide information on the radial evolution of this turbulence. 
The magnetic helicity and the cross-helicity, which express the correlation between the plasma velocity and the magnetic field, are used to characterize the turbulence. Wave number spectra are computed by means of the Taylor hypothesis applied to time resolved single point Voyager 2 measurements. 
The overall picture we get is complex and difficult to interpret. A substantial decrease of the cross-helicity at smaller scales (over 1-3 hours of observation) with increasing heliocentric distance is observed. At 5 AU the only peak in the
probability density of the normalized residual energy is negative, near -0.5. At 29 AU the probability density becomes doubly peaked, with a negative peak at -0.5 and a smaller peak at a positive values of about 0.7. 
A decrease of the cross-helicity for increasing heliocentric distance is observed, together with a reduction of the unbalance toward the magnetic energy of the energy of the fluctuations.
For the smaller scales, at 29 AU the normalized polarization is small and positive on average (about 0.1), but it is zero at 5 AU. For the larger scales, the polarization is low and positive at 5 AU (average around 0.1) while it is negative (around - 0.15) at 29 AU.
% For the smaller scales, we found that at 29 AU  the normalized polarization is small and positive on average  (about 0.1), it is instead zero at 5 AU. For the larger scales, the polarization is low and positive at 5 AU (average around 0.1) while it is negative (around - 0.15) at 29 AU.

% We have investigated the magnetic helicity and the correlation between the plasma velocity and magnetic fields, which is %expressed by the cross-correlation. 
% \dots
% Using data from Voyager 2, we ....

\end{abstract}


\end{frontmatter}


\section{Introduction}
\label{1-introduction}

The solar wind offers a unique scenario where in-situ spacecraft observations can unveil many aspects of very high Reynolds number magnetohydrodynamic turbulence in a magnetized plasma.
One important issue of solar wind studies is to ascertain the dynamics of its fluctuating magnetic and (plasma) velocity fields. These studies involve the measurement and analysis of features which range from large-scale structures originating in the solar corona down to the kinetic scales of the ion Larmor radius. 
Since the earliest observations of the solar wind \cite{mg1982b,bd1971} it has been clear that the solar wind has strong variations in velocity, density, magnetic field and temperature which cannot be described only in terms of the superposition of outward propagating Alfv\'en waves.
The general picture (see, e.g., \cite{Bruno2013}) is that solar wind fluctuations are generated at the solar surface, where new structures continuously appear and fade away, and then propagate outwards through large wavelength Alfv\'en waves. Then energy is gradually transferred toward smaller scales by nonlinear processes until it goes into plasma particles in the form of heat. It is believed that stream-shear instabilities play a major role in the triggering of this transfer process.

The observation of the variation of the solar wind parameters with the distance from the sun is essential for understanding the evolution of the interplanetary fluctuations and has been made possible by the launch of spacecraft.
In particular, the data from the Voyager mission offer a good opportunity to obtain spectral information and investigate the structure of solar wind and its driving mechanisms (underlaying dynamics) close to the solar equatorial plane over a large range of heliocentric distances and solar wind conditions and also beyond the termination shock and towards the edge of the heliosphere.

% The data from the Voyager mission allow to obtain spectral information of the plasma evolution and these information can be related to the macroscopic structure. From the spectra of invariants one can learn much about the state and dynamics of the solar wind. The most important are the total energy per unit mass, the magnetic helicity and the cross helicity.
% % This paper will investigate the magnetohydrodynamics turbulence by means of the statistical properties of its magnetic helicity and cross-helicity densities. 

In this paper we characterize the dynamical structure of the solar wind magnetohydrodynamic turbulence at distances of 5 and 29 astronomical units (AU) from the Sun in terms of both the distribution of cross-helicity and magnetic helicity using data from the Voyager 2 mission. The spectral analysis of these quantities provides information on the plasma evolution over a large range of scales which, in the end, describe the overall macroscopic state of the solar wind.
The magnetic helicity and the cross-helicity both are important characteristics of the solar wind turbulence.
% whose dynamics, as opposed to a neutral flow,
%is determined by the interplay of two dynamic fields, the velocity and the magnetic field.
Like the kinetic helicity and the total energy, both quantities are invariants of the flow when diffusivity and resistivity are not taken into account. 
% In fact, as opposed to neutral fluid flows, the evolution of a magnetohydrodynamic system is determined by the interplay of two dynamic fields, velocity and magnetic field. This carries in new variables to describe the state of the fluid, as the magnetic helicity and the cross-helicity, which, as the kinetic helicity and the energy, are invariants of the flow when diffusivity and resistivity are not taken into account.
A non-zero magnetic helicity indicates the lack of reflectional symmetry in the flow and is related to the dynamo effect. The cross-helicity is proportional to the correlation between velocity and magnetic field fluctuations and measures the relative importance of Alv\'en waves in the global fluctuation. As many studies in literature have shown (see, e.g. \cite{Tu1995,yokoi2013}), the normalized cross-helicity measures the coupling of the magnetic and plasma fields and thus is correlated to the self-production of turbulence.
%significantly influences the turbulence of the flow, contributing to the dynamo effect and saturating the forward cascade processes. % \cite{2014}.
%  Such variables 
Many studies in literature (see, e.g. \cite{Tu1995,gr1995}) indicate that the solar wind turbulence evolves by gradually reducing these correlations and by showing a predominance of the energy associated to the magnetic 
fluctuations with respect to that of the velocity fluctuations.
%component.

%%%%%%%%%%%%%%%%%%%%%%%%%%%%%%%%%%%%%%%%%%%%%%%%%%%%%%%%%%%%%%%%%%%%%%%%%

% Citare: Goldstein-Matthaeus review 1995 \cite{gr1995}, , \cite{demoulin2002}, Roberts-Klein-Mattaeus review 1987 osservazioni Voyager \cite{rkgm1987}, Yoshino-Hokoi \cite{yh2011}.

The definition and the significance of the magnetic helicity and cross-helicity are reviewed in section \ref{2-helicity}. In section \ref{3-results} we present the results obtained from the Voyager 2 time series using reconstruction methods for spectral spectral analysis of lacunous data previously tested on \cite{tmb2014}. % on Voyager 2 data.

% Demouilin 2006 \cite{dpb2006}

%%%%%%%%%%%%%%%%%%%%%%%%%%%%%%%%%%%%%%%%%%%%%%%%%%%%%%%%%%%%%%%%%%%%%%%%%
\section{Helicity and cross-helicity}
\label{2-helicity}
\begin{figure}
 \centering
 \includegraphics[width=.76\textwidth]{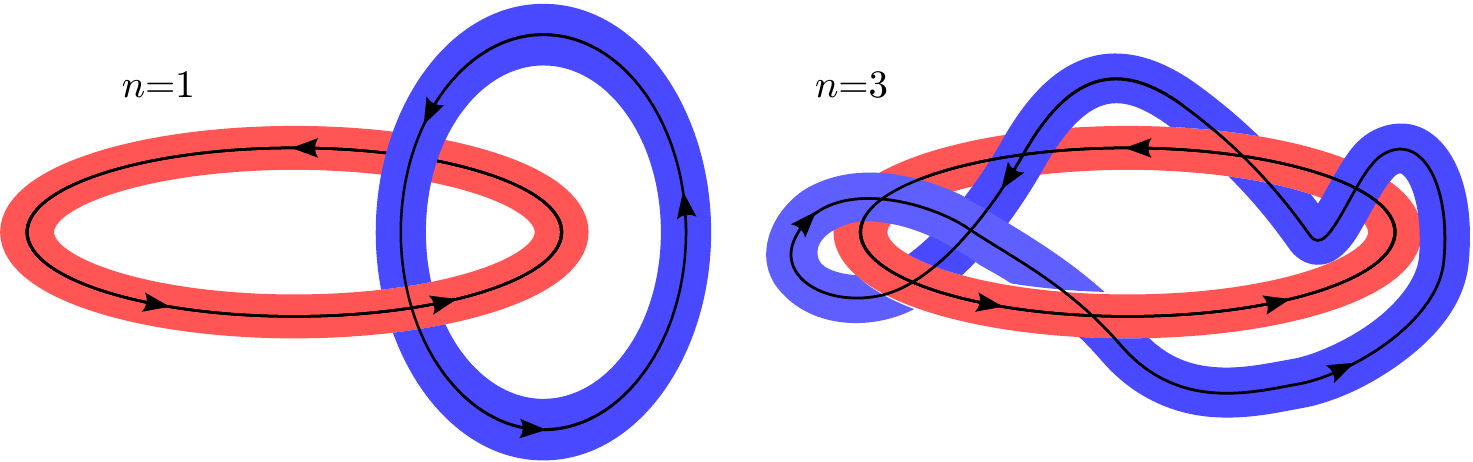}
 \caption{
 \textcolor{black}{Sketch of a configuration of linked vortex/magnetic tubes with different number of windings $n$, that produce a non-zero helicity. Kinetic helicity: red and blue tubes are vorticity tubes. Magnetic helicity: red and blue tubes are magnetic tubes. Cross-Helicity: the red tube is a vortex tube and the blue tube is a magnetic tube (or the opposite).}
 }
 \label{fig.schema}
\end{figure}

%%%%%%%%%%%%%%%%%%%%%%%%%%%%%%%%%%%%%%%%%%%%%%%%%%%%%%%%%%%%%%%%%%%%%%%%%
Since the seminal paper by Moffatt \cite{moffatt1969}, the concept of helicity has been used as a tool to describe the structure of a fluid or plasma flow. From a mathematical point of view, the helicity associated with a vector field $\ug$ in a domain $V$ (bounded or unbounded) is defined as the integral of the scalar (dot) product of the vector and its curl, that is $\int_V \ug\cdot(\nabla\wedge\ug)\,{\rm d}V$. In a plasma flow, we can define the kinetic helicity $H_k$ associated with the velocity field $\ug$ and its curl, the vorticity field $\omegag$, and the magnetic helicity $H_m$ associated with the magnetic potential $\Ag$ and its curl, the magnetic field $\Bg$:
% The helicity associated to a fluid flow in a domain $V$, which can be either bounded or unbounded, is defined as the integral of the scalar product between the velocity field $\ug$ and its, curl, the vorticity field $\omegag$:
\begin{equation}
 H_k = \int_V \ug\cdot\omegag\,{\rm d}V.
\label{def.hk}
\end{equation}
\begin{equation}
 H_m = \int_V \Ag\cdot\Bg\,{\rm d}V.
\label{def.hm}
\end{equation}
The quantities $h_k=\ug\cdot\omegag$ and $h_m=\Ag\cdot\Bg$ represent the kinetic and magnetic helicity density within the flow domain.
Both the helicities, $H_k$ and $H_m$, as well their densities, $h_k$ and $h_m$, are pseudoscalar quantities, that is, they change sign switching from a right-handed to a left-handed frame of reference. 
% In the following, we will always use a right-handed Cartesian frame.
% The helicity has been introduced for the first time by ...
Although the origin of the concept of helicity can be traced to the works by Kelvin \cite{kelvin1869} on the dynamics of vorticity in an inviscid flow, and by Woltjer \cite{w1958a} on ideal resistiveless magnetohydrodynamics, it was first explicitly introduced in more recent times by Moffatt in 1969 \cite{moffatt1969}.
Its name is derived from the fact that, for a given enstrophy $\mid\!\omegag\!\mid^2$, helicity is maximum when the vorticity $\omegag$ is parallel to the velocity $\ug$, a situation in which the streamlines are locally right-handed (positive $h_k$) or left-handed (negative $h_k$) helices about the $\omegag$ axis with a pitch proportional to $h_k/\omega^2$.
Helicity is related to the flow kinematics since it gives a measure of the linkage of vortex lines of the flow. By assuming, for example, that the vorticity is zero in the whole domain except inside two closed and untwisted but reciprocally intertangled vortex tubes, it can be shown (see, e.g., \cite{bf1984,mt1992,pevtsov2008}) that the helicity is equal to $\pm 2n\Gamma_1\Gamma_2$, where $\Gamma_1$ and $\Gamma_2$ are the circulations of the two vortex tubes (vorticity flux) and $n$ is the number of windings of the two tubes \textcolor{black}{(see figure \ref{fig.schema})}.
More generally, \cite{bf1984,cb2014,dpb2006,moffatt2014}, if we assume that the vorticity (or the magnetic field) is concentrated in a finite number of closed tubes, the total helicity can be written as the sum of the contribution of the helicities of the individual tubes, coming from their twisting, and of a contribution of their mutual helicity due to their winding number, a measure of the winding of the tubes about each other.
Although, in general, a vorticity or magnetic potential distribution cannot be simply decomposed into a finite number of simple non-overlapping flux tubes, this simple sketch helps to understand the helicity kinematic significance.

In the same way as the kinetic helicity is a material invariant of the flow in an inviscid (neutral) fluid, which is a direct consequence of Kelvin's theorem, the magnetic helicity $H_m$ is an invariant of ideal inviscid and resistiveless magnetohydrodynamic flow as first proven by Woltjer \cite{w1958a}. The proof of this invariance for ideal magnetohydrodynamic equations depends on the boundary conditions for $\ug$ and $\Bg$ and requires vanishing surface flows on the boundaries. Under such conditions, magnetic helicity is one of the three quadratic invariants (often called ``rugged'' invariants), together with the total energy $E=\int_V \left[u^2+B^2/(\mu_0\rho)\right]{\rm d}V$ and the cross-helicity black {$H_c$},
\begin{equation}
 H_c = \int_V \ug\cdot\Bg\,{\rm d}V.
\label{def.crossh}
\end{equation}
From a geometric point of view, the cross-helicity gives a measure of the degree of linkage of the vortex and magnetic flux tubes within the flow. In fact, if one again considers the simplest archetypical sketch, where vorticity and magnetic fields are concentrated in two thin untwisted closed tubes with are intertangled, the cross-helicity is equal to $2n\Gamma\Phi$, where $n$ is the number of linking or winding of the two tubes, and $\Gamma$ and $\Phi$ are the vortex and magnetic flow across the two tubes. Therefore, a normalized cross-helicity gives the measure of the knottedness of the vortex tube(s) with the magnetic tube(s).

While the physical meaning of energy is clear enough to omit further description, a brief discussion on the dynamical importance of $H_m$ and $H_c$ beyond their kinematic description can be helpful. Both quantities $H_m$ and $H_c$, as well their densities, $h_m$ and $h_c=\ug\cdot\Bg$, are pseudoscalar quantities, that is, they change sign under a reflection of the coordinate system. This implies that they are zero in a mirror-symmetric system and non-zero values can appear only when this symmetry is broken in the flow.
Because the value of these inviscid (ideal flow) invariants cannot be modified by nonlinear terms, they constrain the overall dynamics and, therefore, their value and spectral distribution can give valuable information on the dynamics of turbulence in the solar wind.
Of particular interests are the magnetic and cross-helicities coming from plasma and magnetic field fluctuations.
% ****

% 
% Nonvanishing $H_m$ may also be viewed as a measure of the polarization of wavelike magnetic fluid.
% 
% 
% The helicity is an invariant, alongside energy, in an inviscid flow. Kelvin's theorem states that in an ideal fluid vortex lines behave like material lines. As a consequence, we have the invariance of kinetic helicity (cite).
% A similar results hold for ideal magnetohydrodinamics when we consider the magnetic helicity (cite).
% % \begin{equation}
%  H_B = \int_V \Bg\cdot\Ag\,{\rm d}V.
% \end{equation}

The magnetic helicity plays a central role in the dynamo effect, the so-called $\alpha$-dynamo. In fact, the presence of a magnetic field without reflectional symmetry, that is, with a non-zero magnetic helicity, generates fluctuations of the magnetic field, which, to first order approximation, are proportional to the helicity (\cite{mt1992,moffatt1978,yokoi2013}).
Compared to the magnetic energy, which appears to be transferred to the small scale fluctuations, magnetic helicity presents an inverse cascade, that is, it cascades toward the larger scales of the flow (see \cite{amp2007}). As a consequence, magnetic helicity is depleted much more slowly than energy in a freely evolving flow, while in a forced flow small-scale helical forcing can produce large-scale magnetic fields.

The turbulent cross-helicity is directly related to the coupling coefficients for the mean vorticity in the electromotive force and for the mean magnetic-field strain in the Reynolds stress tensor \cite{yokoi2013}. This relationship suggests that the cross-helicity is important where inhomogeneities in the flow and in the magnetic field appear. Since such large-scale structures are ubiquitous in astrophysical phenomena, cross-helicity is expected to play a role in such flows. In the presence of cross-helicity, the mean vortical structures contribute to the electromotive force: the generation of magnetic field due to this effect is called the cross-helicity dynamo (or $\beta$-effect). In fact, in presence of a large scale vortical motion, the turbulent cross-helicity contributes to the electromotive force aligned with the large-scale vorticity. Provided that the velocity and magnetic field fluctuations are correlated, a contribution to the electromotive force parallel (when $h_c>0$) or antiparallel (when $h_c<0$) to the mean vorticity arises \cite{yb2011}. %, proportional to $\langle\ug'\cdot\omegag'\rangle\langle\omegag\rangle$ \cite{yb2011}.
Moreover, more recently direct numerical simulations of magnetohydrodynamic turbulence have shown that, in association with high values of the cross-helicity, a blocking effect on the spectral transfer of energy is observed and results in energy accumulation in the system. This is concomitant with an increase of the exponent of the spectrum with the cross-helicity level. The spectral exponent increases toward the limiting value of 2 \cite{msf2009}. %\cite{breech2005}
% ****

It should be noted that decaying \textcolor{black}{incompressible} magnetohydrodynamic turbulence, in the absence of large-scale mean shear, tends to amplify the cross-helicity due to a phenomenon called ``dynamic alignment'' \cite{bmc2009} which tends to increase the correlation between the velocity and magnetic field by locally aligning the directions of their polarizations while the energy decays, an effect which is stronger at smaller scales. However, observations indicate that this effect does not play a major role in the solar wind, where a gradual reduction of the content of Alfv\'enic (correlated) fluctuations with the heliocentric distance is observed, in particular near the equatorial plane \cite{mmb2004}.
\section{Voyager 2 data analysis}
\label{3-results}

\begin{table}
%  \begin{tabular}{lcccccccc}
%      &   r    & $\langle u\rangle$ & $\langle B\rangle$ & $E_k$   & $E_b$         & $\langle h_m\rangle$          & $\langle h_c\rangle$  \\
%      &   [AU] &  [km/s]            &   [km/s]           & [km$^2$/$s^2$] &[km$^2$/$s^2$] & [km$^3$/$s^2$] &[km$^2$/$s^2$]\\
% \hline
% 1979-DOY 1--180 & $4.5\div4.9$   &$4.542\cdot10^{2}$  &         & $2.57\cdot10^3$ &                  && $15.8$\\
% 1989-DOY 5--100 & $29.2\div29.8$ &$4.931\cdot10^{2}$  &         & $5.863\cdot10^2$& $3.106\cdot10^3$ && $3.037\cdot10^2$              &\\
% \hline
%  \end{tabular}
\centering
\begin{tabular}{llcc}
\toprule
   && 1979      &   1989\\
   && DOY 1-180 & DOY 5-100\\
\midrule
$r$ &[AU]                            & $4.48\div5.28$   & $28.03\div28.94$\\
% $\langle u\rangle$ &  [km/s]         & $4.542\cdot10^{2}$ & $4.931\cdot10^{2}$\\
$\langle u\rangle$ &  [km/s]         & $454 \pm 43$ & $464 \pm 39$\\
$\langle B\rangle$ &  [nT]           & $0.981$            & $0.207$\\
$E_k$              &  [km$^2$/s$^2$] & $1.224\cdot10^{3}$ & $1.005\cdot10^{3}$\\
% $E_b$              &  [km$^2$/s$^2$] & $5.863\cdot10^2$   & $3.106\cdot10^3$\\
$E_b$              &  [km$^2$/s$^2$] & $1.418\cdot10^3$   & $2.341\cdot10^3$\\
% $\langle h'_m\rangle$& [km$^3$/s$^2$] &\\
$H'_c$& [km$^2$/s$^2$] & $1.58\cdot10^1$  &   $-3.04\cdot10^2$\\
% $\langle h'_m\rangle$& [km$^2$/s$^2$] & $2.02\cdot10^5$  &   $-4.55\cdot10^5$\\
$H'_m$& [km$^2$/s$^2$] & $4.16\cdot10^{10}$  &   $-9.79\cdot10^{10}$\\
\bottomrule
\end{tabular}

\caption{Solar wind data: $r$ is the heliocentric distance, $\langle u\rangle$ and $\langle B\rangle$ are the mean values of the modulus of the velocity and the magnetic field, $E_k$ and $E_m$ are the turbulent kinetic and magnetic energy per unit mass, $\langle u'_ju'_j\rangle/2$ and $\langle b'_jb'_j\rangle/2$, and $H'_m$ and $H'_c$ are the mean values of the magnetic helicity and cross-helicity of the fluctuations. All statistics are computed by averaging the time series over the periods indicated, except the magnetic helicity which is obtained by integrating its spectrum.}
 \label{tab.1}
\end{table}

\subsection{Voyager 2 data}

%%%%%%%%%%%%%%%%%%%%%%%%%%%%%%%%%%%%%%%%%%%%%%%%%%%%%%%%%%%%%%%%%%%%%%%%%%%%%%%%%%%%%%%
\begin{figure}
\hspace*{-0.024\textwidth}
\includegraphics[width=0.46\textwidth]{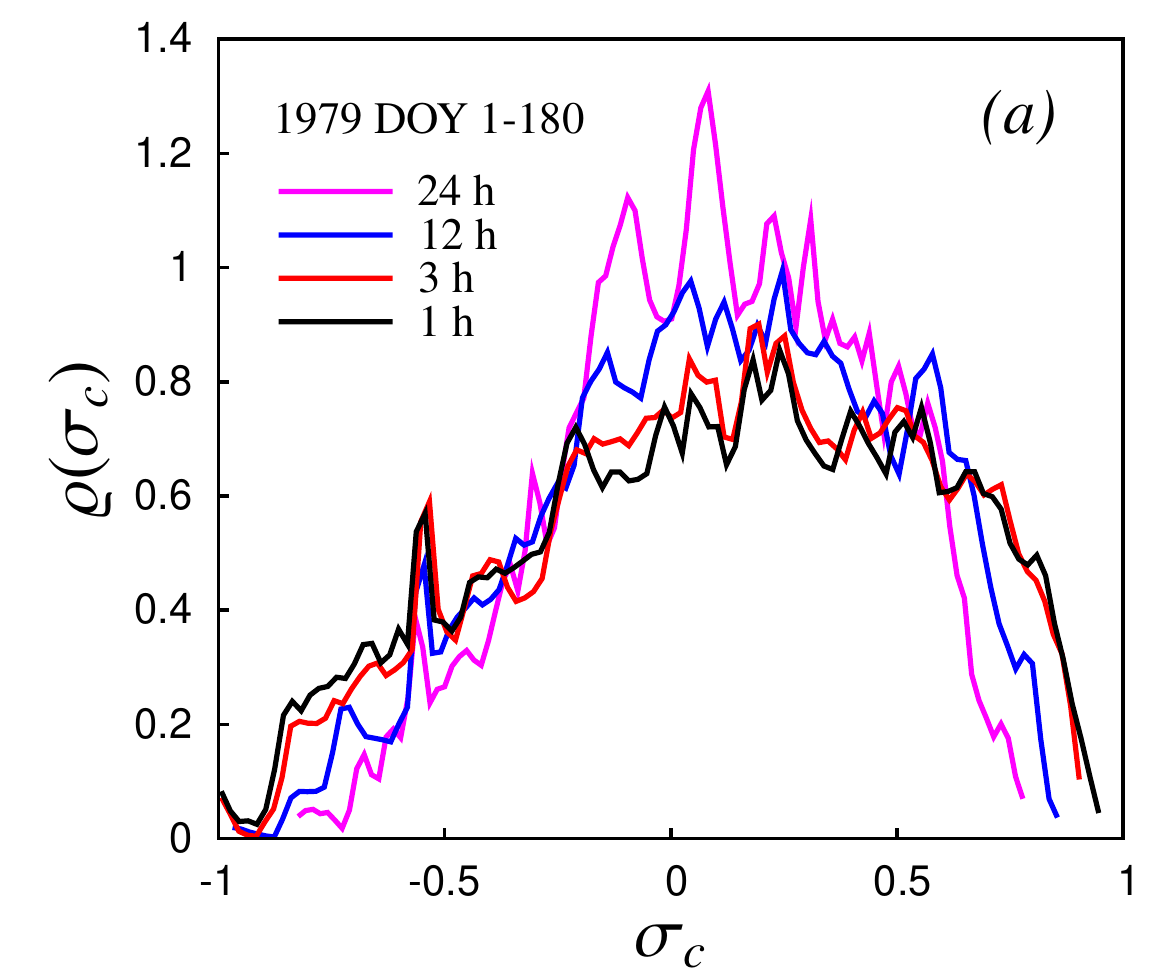}
\hspace*{0.018\textwidth}
\includegraphics[width=0.46\textwidth]{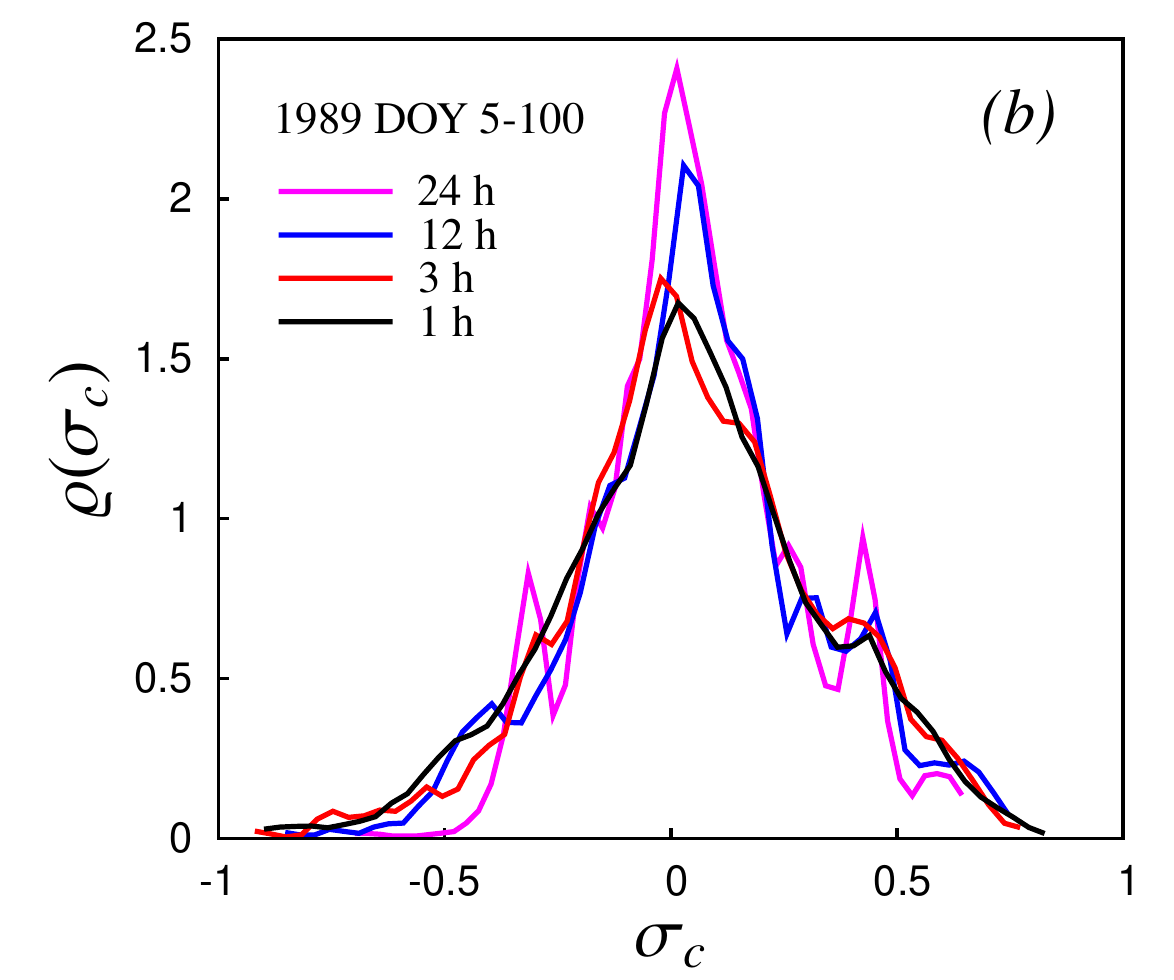}\\
\hspace*{-0.024\textwidth}
\includegraphics[width=0.46\textwidth]{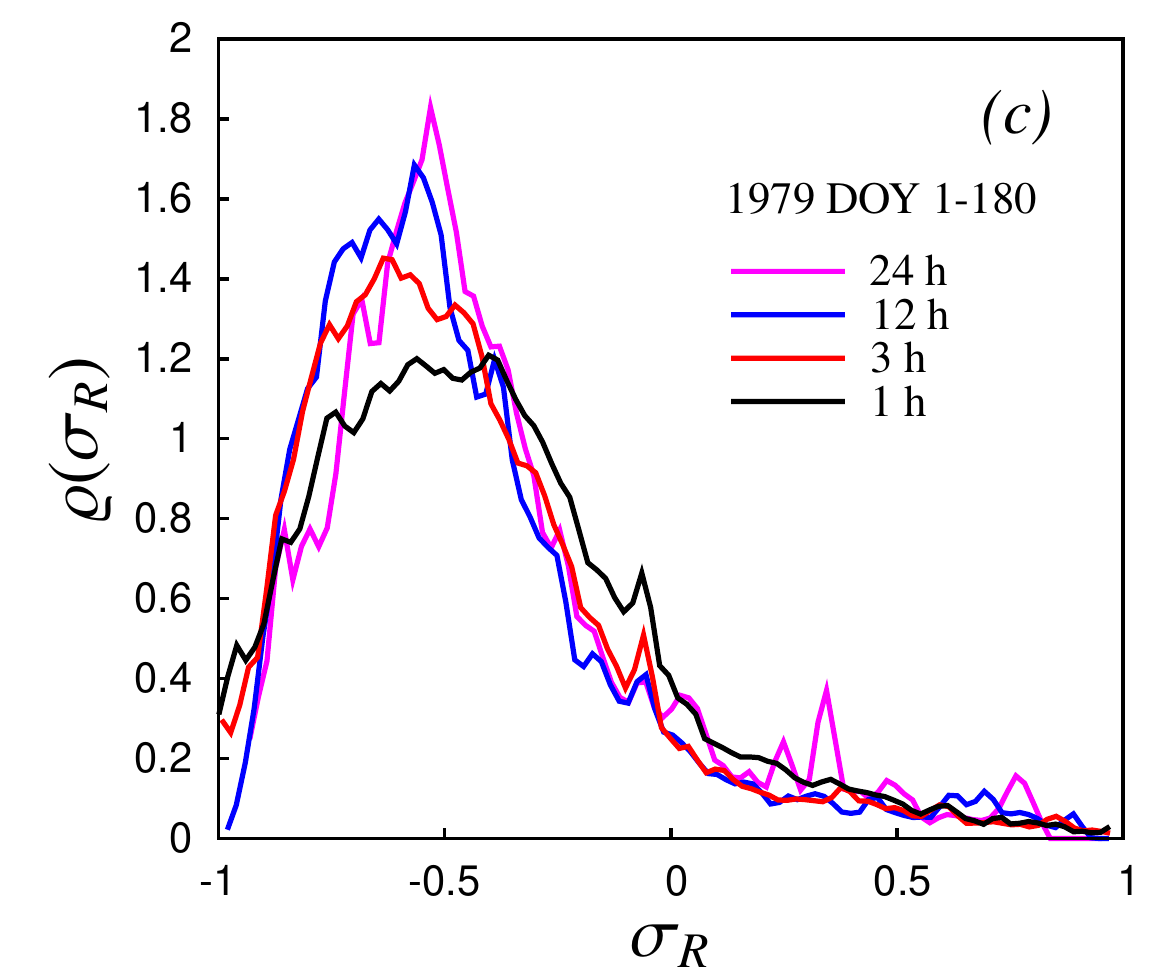}
\hspace*{0.018\textwidth}
\includegraphics[width=0.46\textwidth]{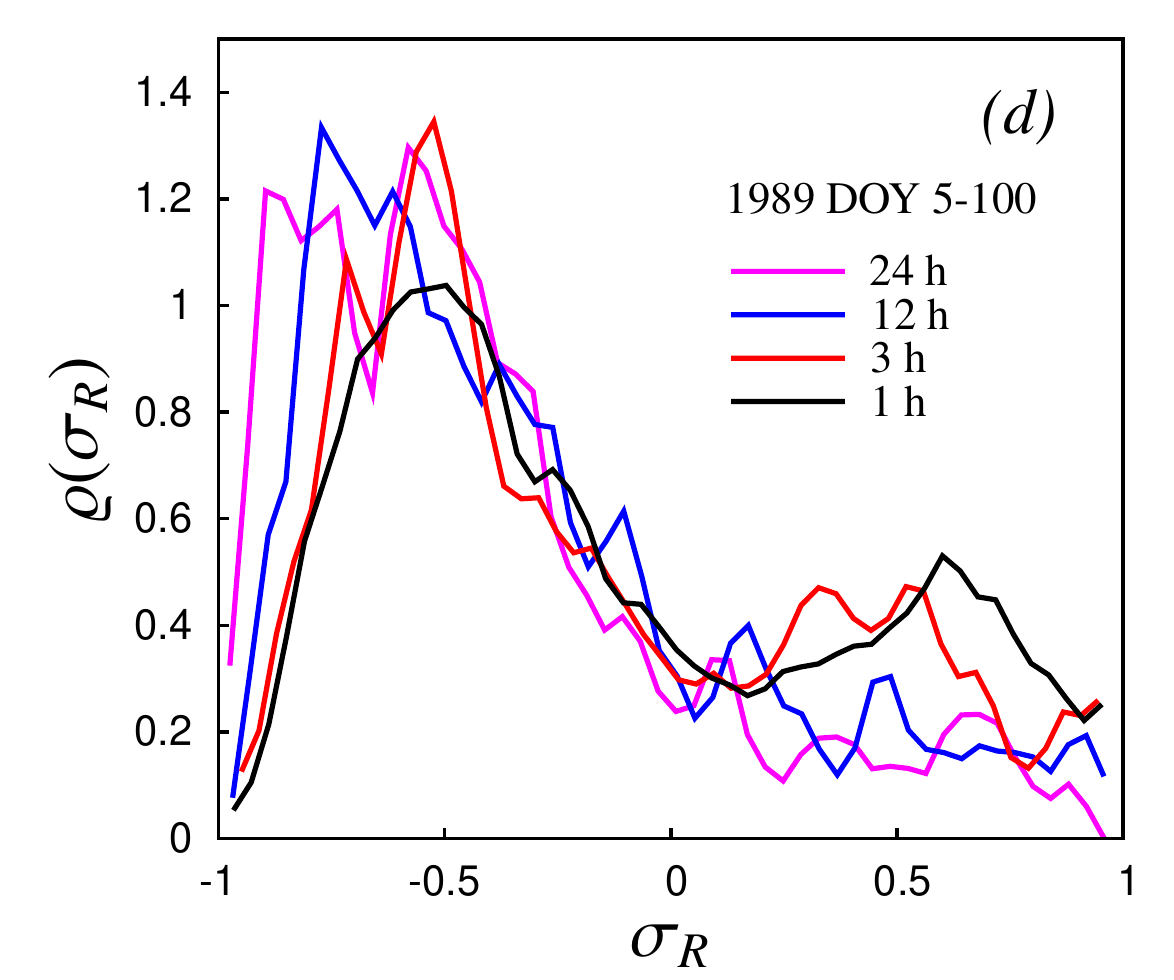}\\
\includegraphics[width=0.499\textwidth]{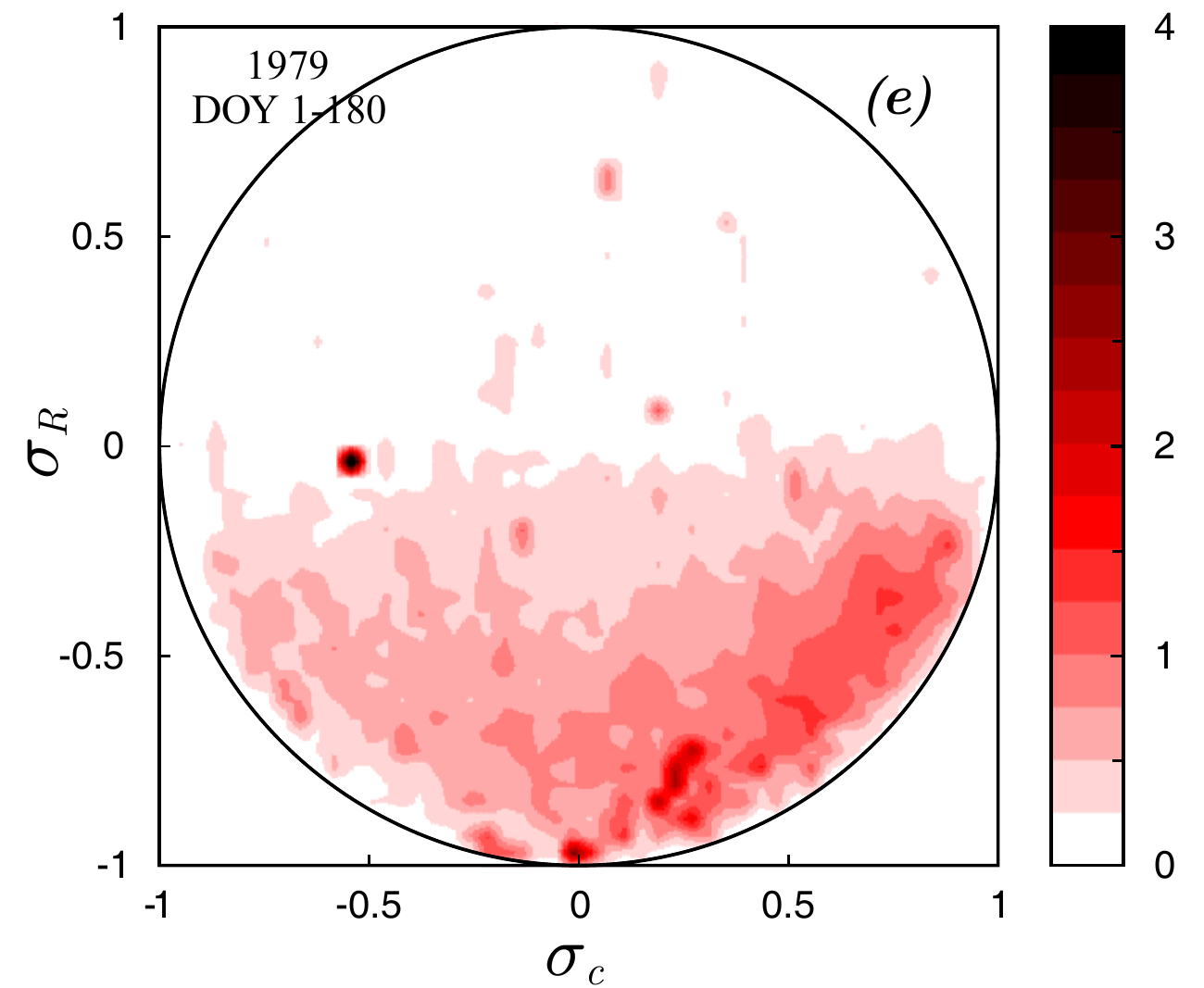}
\includegraphics[width=0.499\textwidth]{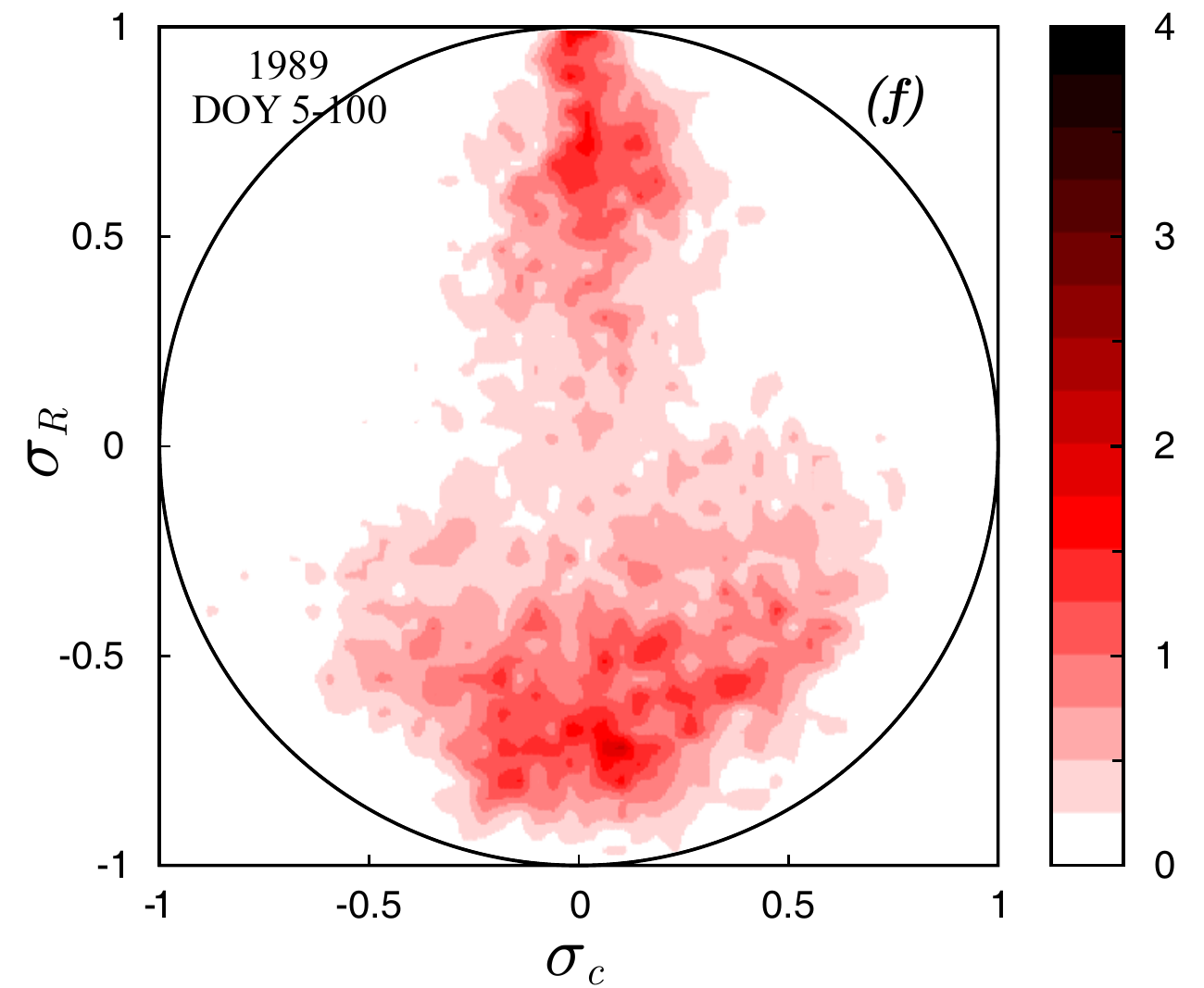}\\
 \caption{Probability density function of the normalized cross-helicity $\varrho(\sigma_c)$  and of the normalized residual energy $\varrho(\sigma_R)$. (a,c) 1979 data DOY 1-180, (b,d) 1989 data, DOY 5-100. The different curves refer to 1, 3, 12 and 24 hour averages. Panels (e) and (f) show the joint probability density function of $\sigma_c$ and $\sigma_R$ computed from hourly averages. The circle indicates the upper bound $\sigma_c^2+\sigma_R^2\le1$.}
 \label{fig.pdf}
\end{figure}
%%%%%%%%%%%%%%%%%%%%%%%%%%%%%%%%%%%%%%%%%%%%%%%%%%%%%%%%%%%%%%%%%%%%%%%%%%%%%%%%%%%%%%%
\begin{table}
 \centering
 %\sigma_c
\begin{tabular}{lcccc}
\toprule
        & \multicolumn{2}{c}{$\sigma_c$}         &\multicolumn{2}{c}{$\sigma_R$}\\
        & 1979               & 1989              & 1979               & 1989\\
\midrule
1 hour  & $9.63\cdot10^{-2}$ & $4.43\cdot10^{-2}$& -0.408             & -0.133            \\
3 hours & $1.07\cdot10^{-1}$ & $5.59\cdot10^{-2}$& -0.459             & -0.194            \\
12 hours& $1.06\cdot10^{-1}$ & $6.09\cdot10^{-2}$& -0.460             & -0.322            \\
24 hours& $9.67\cdot10^{-2}$ & $5.88\cdot10^{-2}$& -0.405             & -0.404            \\
\bottomrule
\end{tabular}
% 
% %\sigma_R
% \begin{tabular}{lcc}
%         & 1979               & 1989\\
% 1 hour  & -0.408             & -0.133            \\
% 3 hours & -0.459             & -0.194            \\
% 12 hours& -0.460             & -0.322            \\
% 24 hours& -0.405             & -0.404            \\
% \end{tabular}

 \caption{Average values of the normalized cross-helicity $\sigma_c$ and of the normalized residual energy $\sigma_R$ of the fluctuations as a function of the averaging time.}
 \label{tab.2}
\end{table}
%%%%%%%%%%%%%%%%%%%%%%%%%%%%%%%%%%%%%%%%%%%%%%%%%%%%%%%%%%%%%%%%%%%%%%%%%%%%%%%%%%%%%%%

%1.I periodi analizzati (descrizione):
The present analysis uses Voyager 2 data from the solar wind plasma and magnetic field experiments to analyse the solar wind turbulence in the outer heliosphere \cite{voyager}. These experiments provide data on the plasma velocity, the density, and the magnetic field. The three components of the magnetic field $\Bg$ and of the plasma velocity $\ug$ are given in the heliocentric spherical RTN coordinate system $\left\{ \eg_R, \eg_T, \eg_N \right\}$ where $\eg_R$ points in the radial direction outward from the Sun, $\eg_T$ is in the plane parallel to the solar equator and positive in the direction of solar rotation, and $\eg_N=\eg_R\wedge\eg_T$ points toward the heliographic north completing the right-handed coordinate system. The $(\eg_R,\eg_T)$ plane is inclined with respect to the equatorial plane by an angle equal to the current latitude of the spacecraft. For our study we have selected two periods: the first six months of 1979 and the first three months of 1989, as shown in table \ref{tab.1}.
%Slow wind is characterized by a mean velocity lower than 500 km/s and a relatively high root mean square of the fluctuations, which can be of the same order of the mean flow, while the magnetic energy is higher [Not in 1989 where ratio is .43] than the kinetic energy (see table \ref{tab.1}), as is shown by the Alfv\'en ratio $E_k/E_m$, which is 0.86 during 1979 and 0.43 during 1989. During the two periods conside, Voyager 2 moves from 4.7 to 5.0 astronomical unit (1979) and from 28.03 to 28.94 AU (1989).
The dataset consists of the velocity and magnetic field components and the density from 1-1-1979 00:00 GMT to 29-5-1979 19:00 GMT for a total of about 180 days and from 5-1-1989 to 11-4-1989 for a total of 96 days.
During 1979, the plasma velocity was sampled each 96 s and the magnetic field data are 48 s averages, while during 1989 the plasma data were sampled each 192 s and the magnetic field data are 48 s averages.
These two periods precede the Voyager 2 encounters with Jupiter and Neptune in July 1979 and August 1989. The Voyager 2 heliocentric distance varies from 4.48 to 5.28 astronomical units (1979) and from 28.03 to 28.94 AU (1989) and the solar wind shows a good degree of homogeneity. %Both periods are characterized by a slow solar wind and show a show a good flow homogeneity.

During both periods the solar wind is in the so-called ''slow wind`` state, with a mean velocity of less than 500 km/s and a fluctuations root mean square of about 25 km/s. The slow wind condition is characterized by relatively high fluctuations, with a high fraction of the total fluctuation energy per unit mass in the magnetic energy, which accounts between 54\%\ and 70\%\ of the energy of the fluctuations. During these two periods, the mean velocity is almost aligned with the radial direction, while the magnetic field tends to have a strong component normal to the radial direction.

%2.che cosa facciamo
We characterize the structure of the solar wind in terms of its cross-helicity and magnetic helicity.
The determination of the kinetic and magnetic helicity, as well as the cross-helicity, requires knowledge of the velocity, vorticity, magnetic field and its potential in the whole flow domain. In the solar wind, however, the magnetic and plasma velocity fields are measurable only at the spacecraft position, like the Voyager mission. This restriction implies that only one-dimensional spectra can be determined by using the Taylor hypothesis of frozen flow.
%3. Ricostruzione dati.
However, a major problem in the spectral analysis of the Voyager mission data is the sparsity of the data, which become more and more lacunous as the spacecraft moves away from the Earth. During the 1979 period,  28\%\ of the data are missing for the plasma velocity and 24\%\ for the magnetic field, while during the 1989 period the missing data increase to 61\%\ for the plasma velocity and 60\%\ for the magnetic field. In order to determine spectral quantities it is necessary to use a data reconstruction method. In the present work, we have adopted data interpolation to fill the gaps in the time series.
This method gives correct estimates of the spectral exponents with an error up to 5\%.
\textcolor{black}{The error estimate was done in \cite{tmb2014} on Voyager 2 1979 data. In this work, we use again
Voyager 2 1979 data plus Voyager 2 data from day 5 to day 100 of 1989. We have verified that
in this last case the same estimate is valid.}
%   In this work, we use again Voyager 2 1979 data plus data from day 5 to day 100 of Voyager 2 1989. We have verified that also in this last case the same estimate holds.}
% This method gives correct estimates of the spectral exponents with an error up to 5\%\ as shown in \cite{tmb2014} \textcolor{black}{on Voyager 2 1979 data}.

% The data is missing due to tracking gaps, noise, and other problems.

From the observed time series we define in each time interval the mean and the fluctuating field. Because of the super-Alfv\'enic flow, we can use the Taylor hypothesis of a frozen-in flow. \textcolor{black}{This approximation  is generally accepted in Solar Wind observations (see for instance Matthaeus and Goldstein 1982 \cite{mg1982a} and Roberts and Goldstein 1987 \cite{Roberts1987b}). In particular, a recent paper by Howes, Klein and TenBarge 2014 \cite{hkt2014} shows the validity of the Taylor hypothesis for super-Alfv\'enic flows. This validation is done under the premise that the frequency of the turbulent fluctuations is well characterized by the frequency of the linear waves supported by the Solar wind plasma. As a consequence, it the Taylor hypothesis holds for linear kinetic wave modes in the weakly collisional Solar wind plasma, with the exceptionn of high wavenumber whistler waves. }
%
%, \textcolor{black}{as shown by the analysis by Howes et al. \cite{hkt2014} for solar wind turbulence}.
%
This assumption allows us to transform time into space by the transformation $x = \overline{u} t$ and, consequently, to determine spatial spectra by linking the frequency to the wavenumber, $\kappa= 2\pi f/ \overline{u}$, where $\overline{u}$ is the mean velocity. In the following, we use the Taylor transformation by assuming the mean velocity field as constant in the two spatio-temporal ranges, the first six month of 1979 and the first 3 months of 1989. As can be seen in Table 1, the velocity field was $454 \pm 43$ km/sec in 1979 and  $464 \pm 39$ in 1989. The magnetic field is expressed in Alfv\'en units, that is, in terms of $\bg={\bf B}/(\mu_0\rho)^{1/2}$, where $\rho$ is the plasma density and $\mu_0$ the magnetic permeability of free space. In this way, $\bg$ has the dimension of a velocity and, consequently, the cross-helicity has the same dimensions of an energy per unit mass while the magnetic helicity has the dimension of an energy per unit mass divided by a length.
The magnetic and cross-helicity spectra have been computed from single point measurement by the Voyager 2 spacecraft using data interpolation to reconstruct the missing values (see \cite{tmb2014} for a discussion of the errors introduced by reconstruction methods in the evaluation of the spectra). \textcolor{black}{We applied the formulae first proposed  by Matthaeus et al. \cite{mg1982a,mg1982b} to deduce the magnetic helicity and cross-helicity spectra from the Fourier transforms of the velocity-magnetic field correlations}. For the magnetic helicity of the fluctuations
\begin{equation}
 \hat{h}_m (\kappa) = \frac{2}{\kappa} {\rm Im}\left( S_{TN}(\kappa)\right),
 \label{sp.hm}
\end{equation}
where $S_{TN}$ is the Fourier transform of the correlation between the tangential and normal components of the magnetic field ($
S_{TN}(\kappa)=
\frac{1}{2\pi}
\int
C_{TN}(r){\rm e}^{i\kappa r}
{\rm d}r,
$ with $
C_{TN}(r)=
\left\langle b'_T(x)b'_N(x+r)\right\rangle
\nonumber
$), 
while the cross-helicity of the fluctuations is computed as the Fourier transform of the correlation $u'_R b'_R + u'_T b'_T + u'_N b'_N$ between the velocity and magnetic fields.
% \begin{equation}
%  \hat{h}_c(\kappa) = .
%  \label{sp.hc}
% \end{equation}
% The computational scheme used in this analysis is simple. From the plasma and magnetic field measurements recorded by Voyager 2 (at 100 s and 48 s time resolution) we ... vedi Physica Scripta \cite{tmb2014}....
% Mattheus {\sl et al.} \cite{mg1982a,mg1982b} first showed how to determine the magnetic helicity and its spectrum from single point measurements of the magnetic field. From a single spacecraft or from a limited number of spacecrafts, however, it is not possible to obtain a full three-dimensional spectrum of any variable without resorting to additional hypotheses, this therefore limits any analysis to one-dimensional spectra.
% Helicity has been measublue through the correlation method, by computing the correlation functions $C_{ij}(u,b)$ and the magnetic autocorrelation $C_{ij}(b)$. ... (mettere formule derivazione $h_c$ da correlazione).
Only the assumption of homogeneity is necessary but no assumption of isotropy is made in obtaining the one-dimensional energy and helicity spectra.
% No assumption of isotropy is made in obtaining the one-dimensional energy and helicity spectra.

%%%%%%%%%%%%%%%%%%%%%%%%%%%%%%%%%%
\subsection{Results}
%%%%%%%%%%%%%%%%%%%%%%%%%%%%%%%%%%

\textcolor{black}{In this study the solar wind turbulence is characterized in terms of two parameters, the magnetic helicity $h'_m$ and the cross-helicity $h'_c$ of the fluctuations. 
The last, which is  defined as
$h_c'=\langle\ug'\cdot\bg'\rangle$,  is first considered in the following. The cross-helicity can be rewritten in terms of the normalized cross-helicity $\sigma_c$ }
\begin{equation}
\sigma_c = \frac{2h'_c}{\langle \ug'\cdot\ug'\rangle+\langle \bg'\cdot\bg'\rangle}
\label{def.sigmac}
\end{equation}
and of the residual energy $\sigma_R$
\begin{equation}
\sigma_R = \frac{\langle \ug'\cdot\ug'\rangle - \langle \bg'\cdot\bg'\rangle}{\langle \ug'\cdot\ug'\rangle+\langle \bg'\cdot\bg'\rangle}.
\label{def.sigmar}
\end{equation}
\textcolor{black}{The normalized cross-helicity of the fluctuations, $\sigma_c$, is a function of the correlation between the plasma velocity and magnetic field fluctuations. It becomes equal to the correlation function
$C(\ug',\bg')=\langle\ug'\cdot\bg'\rangle / (\langle \ug'\cdot\ug'\rangle^{1/2}\langle \bg'\cdot\bg'\rangle^{1/2})$ when
the turbulent kinetic energy per unit mass $\langle \ug'\cdot\ug'\rangle/2$ is equal to the magnetic energy of the fluctuation per unit mass $\langle \bg'\cdot\bg'\rangle/2$, a situation which occurs only when the flow is isotropic.
From its definition, $\sigma_c$ is bounded to values between $-1$ and $+1$, which are attained only if the fluctuations of velocity and magnetic field are perfectly correlated, a situation which corresponds to outward propagating Alfv\'en waves ($\sigma_c=1$) and inward Alfv\'en waves ($\sigma_c=-1$).} The \textcolor{black}{imbalance} between the kinetic and magnetic energy can be expressed by the Alfv\'en ratio $r_A=\langle \ug'\cdot\ug'\rangle / \langle \bg'\cdot\bg' \rangle$, that is the ratio between the kinetic and magnetic energy, or by the so-called normalized residual energy \cite{rkgm1987,Roberts1987b}.

% The normalized residual energy, first used by Roberts et al (1987, citare \cite{rkgm1987}) gives the balance between kinetic and magnetic energy (rewritten in Alfv\'en units) and it is related to the so-called Alfven ratio $\lbrace u^2\rbrace / \lbrace u^2\rbrace$. 
The absence of magnetic fluctuations corresponds to a normalized residual energy equal to 1, the absence of kinetic fluctuations corresponds to a normalized residual energy equal to -1, while zero implies equipartition of the energy between kinetic and magnetic fluctuations.
% Beside the cross-helicity, we have also computed the correlation coefficient between the $\ug$ and $\bg$ vectors.
The correlation is linked to the cross-helicity through the residual energy $\sigma_R$ by $C(\ug',\bg')=\sigma_C/(1-\sigma_R^2)^{1/2}$. Therefore, the cross-helicity does not generally equal the cross-correlation and they coincide only if there is equipartition of energy between the two fields.

\begin{figure}
 \centering
 \includegraphics[width=.55\textwidth]{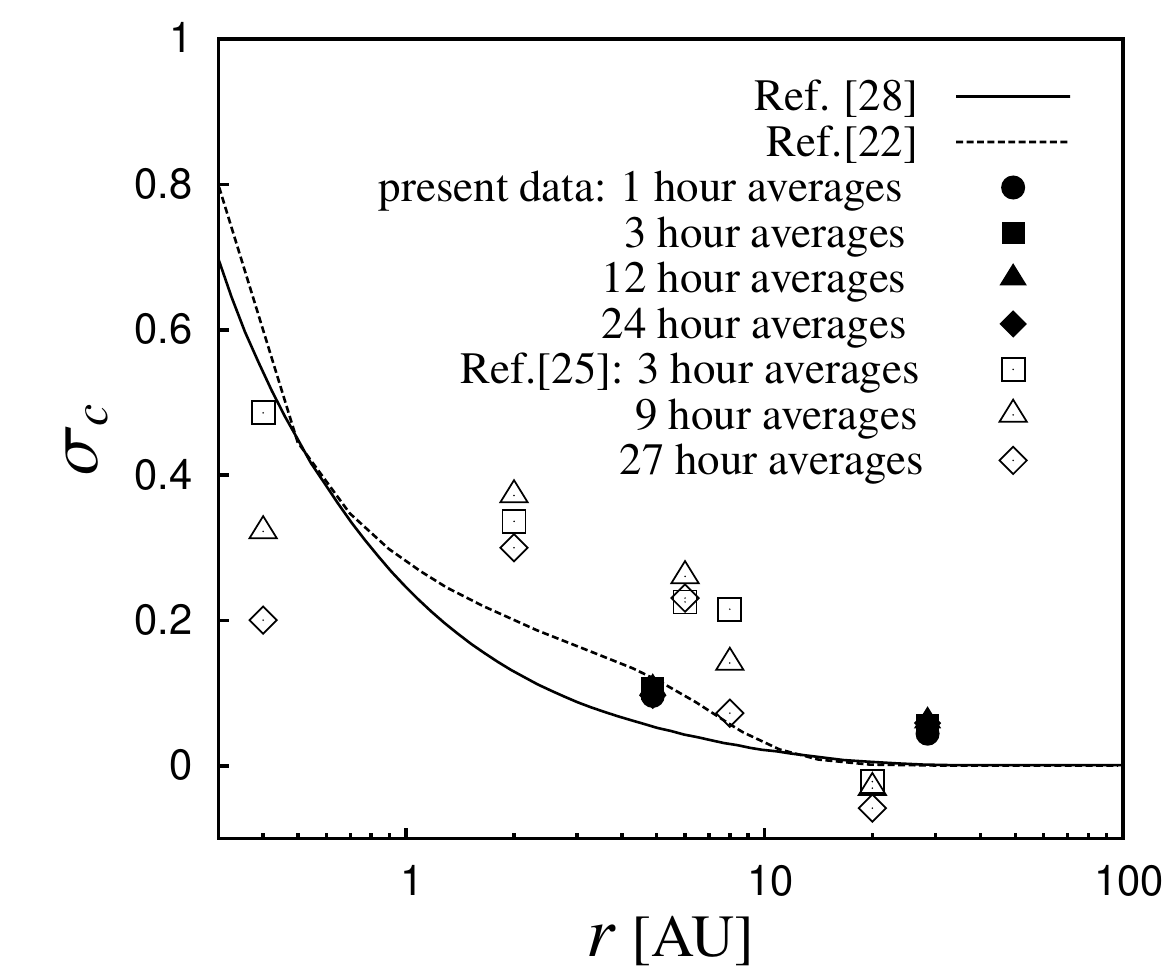}
 \caption{Variation of the cross-helicity with heliocentric distance. Filled symbols show the results of the present analysis of Voyager 2 data (in tabular form in table \ref{tab.2}). The lines represent the mean field models by Breech et al. \cite{breech2005} (continuous line) and by Matthaeus et al. \cite{mmb2004} (dashed line), empty symbols are Voyager and Helios data from \cite{Roberts1987b}.}
 \label{fig.distanza}
\end{figure}

% NORMALIZED CROSS HELICITY

Past analysis of solar wind turbulence in the inner heliosphere \cite{Tu1995,Bruno2013} indicates that the solar wind evolves toward a decreasing correlation between velocity and magnetic field fluctuations and toward an increasing imbalance in favour of the magnetic contribution to the energy of the fluctuations.

Present analysis shows that the greatest imbalance occurs in the less spatially evolved (closest to the Sun) 1979 solar wind, while the more evolved (furthest from the Sun) 1989 solar wind shows a tendency toward energy equipartition, at least at the small scale level. In fact, while the 1989 24-hour average data show a residual energy $\sigma_R$ of about -0.4, the hourly average data indicates a drop to -0.13 (see table \ref{tab.2}). This behaviour was not present in the 1979 data, where all scales had a similar $\sigma_R$, ranging from -0.4 to -0.46. The imbalance in favour of the magnetic energy tends to be less pronounced at small scales. This relative increase in magnetic energy could be due to waves generated by pickup ions, which initially have an unstable ring distribution (see \cite{Smith2006}).

%Smith, C.W., P.A. Isenberg, W.H. Matthaeus, J.D. Richardson. Turbulent heating of the solar wind by newborn interstellar pickup protons. Astrophys J., 638, 508-517, 2006.

Panels (a) and (b) of figure \ref{fig.pdf} show the probability density function of $\sigma_c$ computed using different averaging times in (\ref{def.sigmac}), from 1 hours to 24 hours.
\textcolor{black}{The probability densities have been deduced from the numerical derivative of the empirical cumulative distribution functions \cite{ww1978,libropdf}.}
%, computed in 100 points beween the minimum and the maximum of each variable.
Distributions become more symmetric with increasing heliocentric distance, with no qualitative difference between the different averaging times apart from a small reduction of the variance of $\sigma_c$ with increasing time scale while the mean values remain almost constant (see table \ref{tab.2}). The most significant result is the strong reduction of the normalized cross-correlation variance in the 1989 data, which indicates a marked reduction in the presence of Alfv\'enic fluctuations. The $\sigma_c$ values tend to a symmetric distribution centered around zero. An increasing variance at small scales is observed in 1979, see panel (b) in figure \ref{fig.pdf};  an increasing variance is a common characteristic of the inner heliosphere \cite{Roberts1987b} where broader probability density functions are usually  observed for averages below one day.
% In figure \ref{fig.pdf}, parts (c) and (d), the probability density function of the residual energy $\sigma_R$ is shown. Positive values are almost completely absent during 1979 at 5 AU, with a peak around -0.5 which sharpens with increasing averaging scale. On the contrary, the 1989 distributions at 29 AU are more scatteblue with a non negligible presence of data with positive $\sigma_R$, in particular for shorter averaging time where a secondary peak around 0.5 is clearly visible. Both positive and negative $\sigma_R$ are not correlated with a particular value of $\sigma_c$, which is symmetrically distributed for all values of $\sigma_R$ (see part (f)).
Panels (c) and (d) of figure \ref{fig.pdf} show the probability density function of the residual energy $\sigma_R$. Positive values are almost completely absent during 1979 at 5 AU, with a peak around -0.5 which sharpens with increasing averaging scale. On the contrary, the 1989 distributions at 29 AU are more scattered with a non-negligible data with positive $\sigma_R$, in particular for the shorter averaging times, one and three hours, where a secondary peak around 0.5 is clearly visible. Both positive and negative $\sigma_R$ are not correlated with a particular value of $\sigma_c$, which remains symmetrically distributed around zero for all values of $\sigma_R$ (see the joint probability distribution in panel (f)).

The overall evolution of $\sigma_c$ with heliocentric distance is summarized in table \ref{tab.2} and in figure \ref{fig.distanza}. Moving from 5 to 29 AU close to the ecliptic plane, our data show a moderate decline of $\sigma_c$ which follows well the trend predicted by previous investigations \cite{breech2005,mmb2004} based on mean field Reynolds Averaged MHD simulations where turbulent trasport was taken into account. \textcolor{black}{It should be noted that time variations due to the solar wind activity should not have a major influence on present results because both periods, 1979 DOY 1 - 180 and 1989 DOY 5 - 100, show a slow wind and the time spans are longer than the solar rotation, about 25 days at the equatorial plane. In figure \ref{fig.distanza}, we show also the normalized cross-helicity computed by Robert et al. in 1987 \cite{Roberts1987b} (data from Helios 1 at 0.3 AU and Voyager 2 1985 from 2 to 20 AU), see open symbols. These last computations agree to a lesser extent to the above mentioned Reynolds averaged simulations. However, if one interpolates  data with a same average time interval, one sees that they follow better the older 2004 version of the model \cite{mmb2004} than the updated 2005 version of the same model \cite{breech2005}. In the last case, the transport of cross helicity and radial evolution of Alfv\'enicity was computed with a slightly lower  value of the K\'arm\'an-Taylor constant and slightly different boundary condition at 0.3 AU (similarity lenght of 0.01 AU against 0.025, fluctuation energy of 2200 instead of 2000 km$^2$/s$^2$ and a slightly higher normalized cross-helicity). It should be noted that in \cite{Roberts1987b} the data were not all properly sorted, for example by wind speed. }

%do not agree well with the global evolutive models as for instance those by Breech et al, 2005, and Matthaeus et al., 2004, \cite{breech2005,mmb2004}. 
%In fact, these models do not yet consider the effect associated to the presence of high speed streams, stream shear and magnetic sector crossing \cite{Roberts1987b,Bruno2013}.} 
\textcolor{black}{As the plasma moves away from the Sun, the velocity and magnetic fields become less correlated. This decrease in correlation is in contrast with results obtained by means of numerical simulations of homogeneous and isotropic {\itshape incompressible} magnetohydrodynamic turbulence (e.g. \cite{MCB2006} ) and by phenomenological interpretation carried out on the same system configuration (\cite{bmc2009,bold2005}).
%, that suggests that  dynamic alignment shows up spontaneously as that the flow evolves which induces an intense cross-helicity \cite{bmc2009}, a fact that was used to explain the high correlation between the velocity and the magnetic field in early observations in the innermost Heliosphere \cite{dobro1980}. 
In short, the relevance of the dynamic alignment in {\it compressible} MHD turbulent flows, and in particular in the the Solar wind, is still debated (see, e.g., \cite{Bruno2013} and \cite{MHD} and the references therein). Compressibility, velocity shears and density gradients (and related turbulent diffusion) present in the Solar Wind may overwhelm the effect of the dynamic alignment, driving the turbulence toward lower levels of cross-helicity.}
% In fact, the Solar wind includes many features which can contribute to the depletion of this correlation, which are particularly relevant in the slow wind near the ecliptic, like the compressibility and the presence of large-scale shear and density gradients. These phenomena can overwhelm the effect of the dynamic alignment, driving the Solar wind toward lower cross-helicity.}

%Kraichnan \cite{kraichnan1965} predicted that in isotropic magnetohydrodynamic turbulence kinetic and magnetic energy should be equal on average in the inertial range. As suggested by the normalized residual energy, the smaller scales we can observe could be evolving toward local isotropy.
%A measure of this is the Alfv\'en ratio and the normalized residual energy $\sigma_R$. The data plotted in figure \ref{fig.pdf} show that the smaller scales of the solar wind could have evolved toward local isotropy.

%%%%%%%%%%%%%%%%%%%%%%SPETTRI%%%%%%%%%%%%%%%%%%%%%%%%%%%%%%%%%%%%%%%%%%%%
\begin{figure}
 \includegraphics[width=0.499\textwidth]{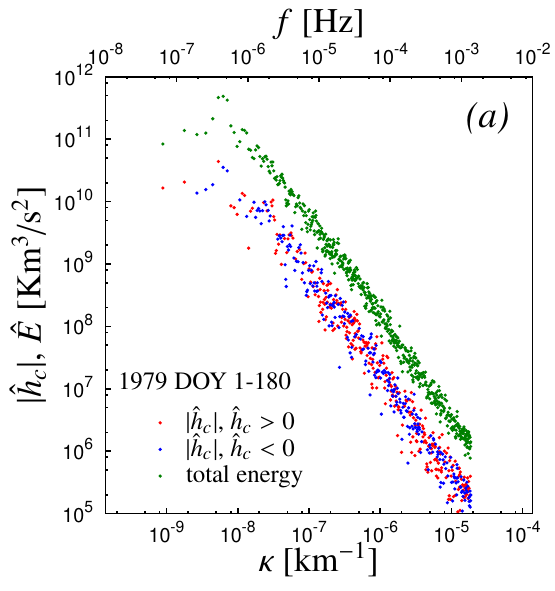}
 \includegraphics[width=0.499\textwidth]{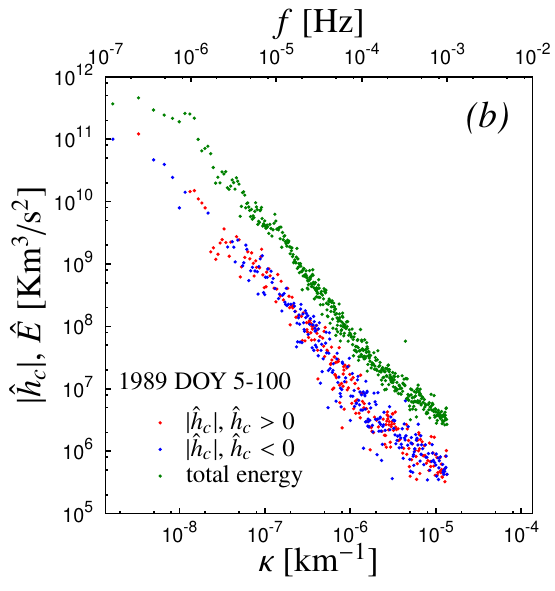}
 \caption{Wave numbers  spectra of the cross-helicity density and of the total energy. These spectra are obtained by means of the Taylor hypothesis, see section \ref{3-results}. The mean speed is that computed over the whole time interval, see Table 1. For clarity, in panel (a) not all the points of the spectra are shown for frequencies above $10^{-6}$ Hz. The same in panel (b) above $10^{-5}$ Hz. 
The small peaks at a frequency of $2.6\cdot10^{−4}$ Hz are likely instrument-related, since such frequencies correspond to about four time the sampling period of the plasma velocity.}
 \label{fig.spettri_hc}
\end{figure}

\begin{figure}
 \includegraphics[width=0.499\textwidth]{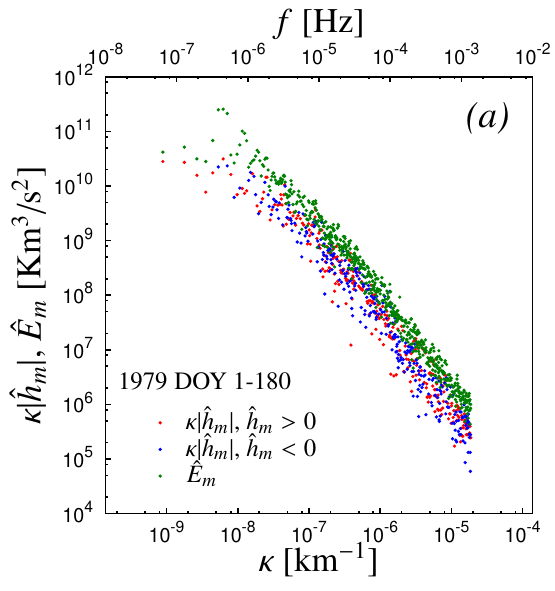}
 \includegraphics[width=0.499\textwidth]{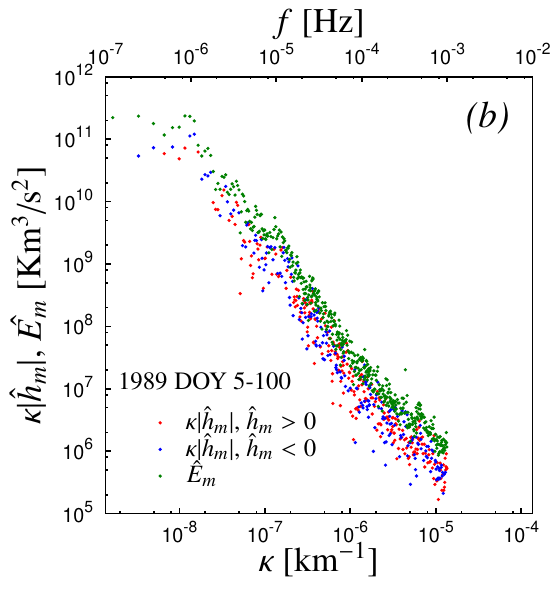}\\
 \includegraphics[width=0.499\textwidth]{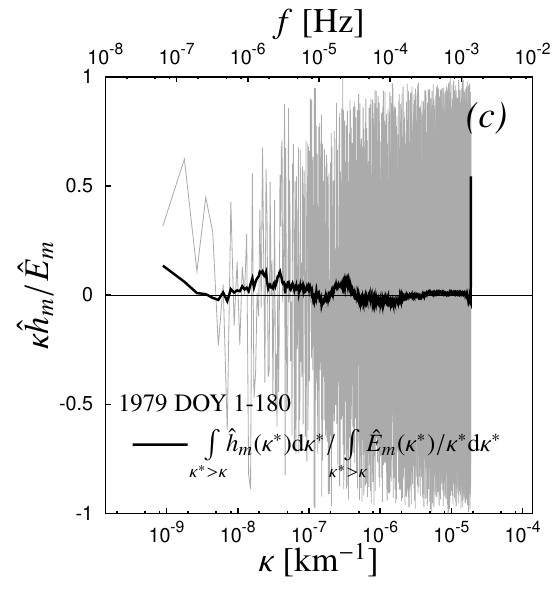}
 \includegraphics[width=0.499\textwidth]{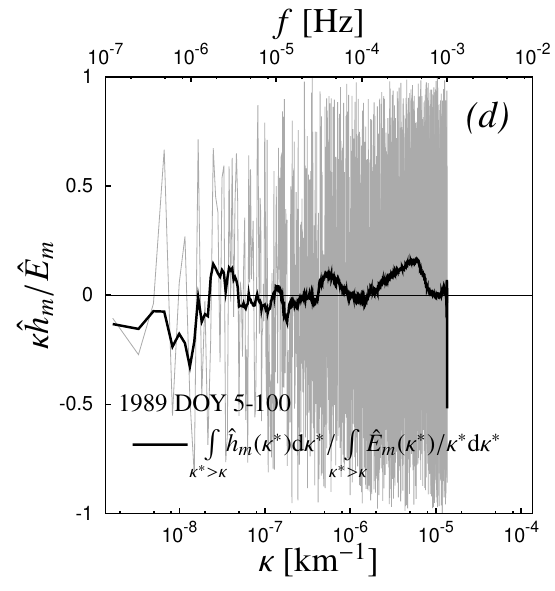}
 \caption{(a,b) Wave numbers  spectra of the magnetic helicity density and of the magnetic energy $\left<b'_jb'_j\right>/2$ in Alfv\'en units. These spectra are obtained by means of the Taylor hypothesis, see section \ref{3-results}. The mean speed is that computed over the whole time interval, see Table 1. For graphic clarity, in panel (a), for frequencies above $10^{-6}$ Hz, not all the points of the spectra of the magnetic helicity are shown. The same in panel (b) above $10^{-5}$ Hz. (c,d) Relative normalized magnetic helicity $\sigma_m(\kappa)=\kappa\hat{h}_m/E_m(\kappa)$. The thick line indicates the cumulative relative helicity integrated from a given 
$\kappa$ to the highest observed wavenumber.}
 \label{fig.hm}
\end{figure}
%%%%%%%%%%%%%%%%%FINE%SPETTRI%%%%%%%%%%%%%%%%%%%%%%%%%%%%%%%%%%%%%%%%%%%%

A closer look at the scale dependence of the cross-helicity can be obtained by looking at its one-dimensional power spectrum. The cross-helicity spectrum can take both positive and negative values. The sign of the cross-helicity spectrum often indicates the propagation of Alfv\'enic waves outwards from the sun (positive helicity)  \cite{mg1982a,mg1982b}, as also contemplated by Belcher \&\ Davis \cite{bd1971}, but periods with mixed helicity sign can appear. 
In our data, see figure \ref{fig.spettri_hc}, there is no preferred sign in any part of the cross-helicity spectra, whose modulus shows a fair power law scaling with an exponent equal to -1.59, close to the spectral exponent of the total energy spectrum, -1.62 during 1979. In the later period, 1989, the spectra become steeper,  and the spectral exponents are -1.74 for the cross-helicity and -1.80 for the total energy.% without any other change in the overall picture.
% We have computed the cross-helicity from the matrix of two-point correlations as in \cite{mg1982a,mg1982b}
This is clearly visible in figure \ref{fig.spettri_hc}, which shows the cross-helicity spectrum compared with the total energy spectrum, the sum of the kinetic and the magnetic energy of the fluctuations.
The number of frequencies with positive and negative values is comparable at all scales, and they balance: the integral cross-helicity can be estimated as 15.8 km$^2$/s$^2$ during 1979 and 304 km$^2$/s$^2$ during 1989, much lower than the total energy of 2640 and 2346 km$^2$/s$^2$ in 1979 and 1989, respectively. Globally the normalized cross-correlation is less than 1\%\ at 5 AU and of the order of 10 percent \textcolor{black}{at 29 AU} and therefore the two fluctuationg fields are very mildly correlated.
%Only a limited set of frequencies present large values of normalized cross-helicity, a sign of the residual presence of \textcolor{black}{the} contribution of Alfv\'enic waves.

Similar to the spectrum of the cross-helicity, the spectrum of the magnetic helicity, which has also a power law range with exponent -2.66  in 1979 and -2.92 in 1989, see panels (a) and (b) in figure \ref{fig.hm}, shows almost an equipartition positive and negative values throughout the inertial range of the spectrum. The exponents of the magnetic energy spectrum in the same range of frequencies are -1.63 and -1.80, respectively (notice that in the ordinate the quantity $\kappa \hat{h}_m$ is represented). This result implies that the relative magnetic helicity of each scale, given by the ratio $\kappa \hat{h}_m(\kappa)/\hat{E}_m(\kappa)$, slowly decreases as $\kappa^{-0.03}$ (5 AU) and $\kappa^{-0.12}$ (29 AU) in the observed wavenumber range. The sign of the  $[\kappa, \infty]$ averaged magnetic helicity, $H_m=\int_\kappa^{+\infty} \hat{h}_m(\kappa^*){\rm d}\kappa^*$ carries the information on the polarization at the different scales. For instance, the sign of first the points of the dark curves in panels in panels (c) and (d) gives the global polarization because of course the largest scales contribution prevails energetically.  
%{fig.hm}
%Therefore, the relative magnetic helicity of each scale, given by the ratio
%$\kappa H_m(\kappa)/\hat{E}_m(\kappa)$, remains rather small in the whole range of wavenumbers we observe, which could not be surprising because we have averaged over a noisy field.
The sign of $H_m/\int_\kappa^{+\infty}(\hat{E}_m/\kappa^*){\rm d}\kappa^*$ is negative at the largest scales during 1989 but is positive during 1979. If we consider all the wave numbers, this value is relatively small: 0.135 at 5 AU and -0.132 at 29 AU. But if we consider the small scale ranges, we see that in 1979 the polarization is negligible, while in 1989 it is definitely positive for the last two decades. It is known that magnetic helicity can change sign and have opposite signs at the largest and smallest scales in magnetohydrodynamic system, an effect of the inverse cascade and its conservation \cite{brandenburg2011}.  
As a general comment, the weak values of the normalized magnetic helicity can be considered a signature of the negligible injection of helicity in the solar wind at the equatorial plane because the presence of the inverse cascade of helicity makes the helicity dissipation much lower than the dissipation of energy, thus the magnetic helicity remain almost constant throughout the system.
%\cite{brandenburg2011}
%can appear beyond 1 AU

% Nota: cross-helicity and energy: forward cascade, magnetic helicity: inverse cascade.

% Bavassano 1998 \cite{bpr1998} -- cross.helicity at high latitudes (confronto) --- evoluzione radiale \cite{mmb2004}.
%%%%%%%%%%%%%%%%%%%%%%%%%%%%%%%%%%%%%%%%%%%%%%%%%%%%%%%%%%%%%%%%%%%%%%%%%
% \begin{figure}
%  \caption{Time series of the magnetic helicity and cross-helicity density: (a) 1979 data, (b) 1989 data.}
%  \label{fig.tempo}
% \end{figure}

%%%%%%%%%%%%%%%%%%%%%%%%%%%%%%%%%%%%%%%%%%%%%%%%%%%%%%%%%%%%%%%%%%%%%%%%%
\section{Conclusion}
\label{4-conclusion}
% ... We found that ...

We use data reconstruction techniques on the Voyager 2 data to
study the radial evolution of the turbulent kinetic and
normalized residual energy of the magnetic helicity and cross-helicity
of the solar wind in ``slow wind'' conditions.
We determine the probability density functions and the spectra of the
magnetic and cross-helicity at 5 and 29 AU from the Sun. We found a
small level of Alfv\'enic fluctuations, which decrease with
heliocentric distance, a possible indication of the evolution toward a
turbulent state dominated by nonlinear interactions. The sign of the
normalized cross-correlation remains positive, which indicates the
slight prevalence of outward propagating perturbation waves.
For the fluctuating field, the different time scales used in the
statistical averages and cross-helicity spectra indicate that the
outward and inward directions of propagation are equally probable in
the whole range of scales we have analyzed. Together with the reduction
of the normalized cross-helicity, this could suggest that, in the
``slow wind'', solar wind turbulence is evolving toward a
developed magnetohydrodynamic 
\textcolor{black}{turbulence. However, data
in 1989 show a sign change of polarization (the small scales are
positively polarized) not present in 1979 data. Also, the polarization
intensity is slightly higher in 1989.  This result is not consistent with
the hypothesized  isotropization of the field predicted by models
of homogeneous and isotropic
magnetohydrodynamic turbulence \cite{kraichnan1965}.
In future work we plan to extend this analysis to the outer part of the
heliosphere and to the compressed solar beyond the termination shock.}

\section*{Acknowledgements}

We wish to ackowledge the support of the Misti Global Seed Funds, MITOR Project ``Laboratory Simulation Of Planet-Solar Wind And Interstellar Medium/Heliosphere Interactions'', 2011-2015.

  \bibliographystyle{elsarticle-num} 
  \bibliography{ejmb_bib}


% \end{thebibliography}
\end{document}